\documentclass[twocolumn,prl,amsmath,amssymb,showpacs,superscriptaddress]{revtex4}
\usepackage{graphicx}
\usepackage{dcolumn}
\usepackage{bm}
\usepackage{latexsym}
\usepackage{amstext}
\usepackage{amsxtra}
\usepackage{color}

\newcommand{\ket}[1]{|#1\rangle}

\begin{document}

\title{Demonstration of a strong Rydberg blockade \\ in three-atom systems with anisotropic interactions}

\author{D. Barredo}
\affiliation{Laboratoire Charles Fabry, Institut d'Optique, CNRS, Univ Paris Sud,\\
2 avenue Augustin Fresnel, 91127 Palaiseau cedex, France }

\author{S. Ravets}
\affiliation{Laboratoire Charles Fabry, Institut d'Optique, CNRS, Univ Paris Sud,\\
2 avenue Augustin Fresnel, 91127 Palaiseau cedex, France }

\author{H. Labuhn}
\affiliation{Laboratoire Charles Fabry, Institut d'Optique, CNRS, Univ Paris Sud,\\
2 avenue Augustin Fresnel, 91127 Palaiseau cedex, France }

\author{L. B\'eguin}
\affiliation{Laboratoire Charles Fabry, Institut d'Optique, CNRS, Univ Paris Sud,\\
2 avenue Augustin Fresnel, 91127 Palaiseau cedex, France }

\author{A. Vernier}
\affiliation{Laboratoire Charles Fabry, Institut d'Optique, CNRS, Univ Paris Sud,\\
2 avenue Augustin Fresnel, 91127 Palaiseau cedex, France }

\author{F. Nogrette}
\affiliation{Laboratoire Charles Fabry, Institut d'Optique, CNRS, Univ Paris Sud,\\
2 avenue Augustin Fresnel, 91127 Palaiseau cedex, France }

\author{T. Lahaye}
\affiliation{Laboratoire Charles Fabry, Institut d'Optique, CNRS, Univ Paris Sud,\\
2 avenue Augustin Fresnel, 91127 Palaiseau cedex, France }

\author{A. Browaeys}
\affiliation{Laboratoire Charles Fabry, Institut d'Optique, CNRS, Univ Paris Sud,\\
2 avenue Augustin Fresnel, 91127 Palaiseau cedex, France }

\date{\today}

\begin{abstract}
We study the Rydberg blockade in a system of three atoms arranged in different 2D geometries (linear and triangular configurations). In the strong blockade regime, we observe high-contrast, coherent collective oscillations of the single excitation probability, and an almost perfect van der Waals blockade. Our data is consistent with a total population in doubly and triply excited states below 2~$\%$. In the partial blockade regime, we directly observe the anisotropy of the van der Waals interactions between $\ket{nD}$ Rydberg states in the triangular configuration. A simple model, that only uses independently measured two-body van der Waals interactions, fully reproduces the dynamics of the system without any adjustable parameter. These results are extremely promising for scalable quantum information processing and quantum simulation with neutral atoms.
\end{abstract}


\maketitle

Engineering quantum many-body systems with a high degree of control and tunable interactions is an active field of research as it is a prerequisite for quantum information processing~\cite{nielsen2000} and quantum simulation~\cite{feynman1982}. Recently,  significant achievements have been obtained towards this goal, e.g. using trapped ions for simulating quantum magnetism \cite{lanyon2011,britton2012,islam2013}. Another platform considered for such tasks consists of systems of neutral Rydberg atoms interacting via the strong and controllable long-range dipole-dipole interaction, which is responsible for the Rydberg blockade~\cite{jaksch2000,lukin2001,comparat2010,weimer2010}. Through this mechanism, multiple excitations with a resonant narrow-band laser are inhibited within a blockade sphere by Rydberg-Rydberg interactions. The dipole blockade provides a way to realize fast quantum gates and to entangle particles, as demonstrated for two atoms \cite{isenhower2010,wilk2010}. This mechanism can in principle be extended  to an ensemble of $N$ atoms, with fascinating applications in quantum state engineering~\cite{saffman2010}.

Although the picture of a blockade sphere has been remarkably successful at describing many recent experiments~\cite{singer2004,heidemann2007,pritchard2010,dudin2012,dudin2012b,schauss2012,urban2009,gaetan2009,gunter2013,bidermann2014}, some theoretical works question this simple approach. Even for the case of $N=3$, some situations have been identified where nearly resonant dipole-dipole interactions \cite{pohl2009}, the non-additivity of the van der Waals potentials \cite{cano2012}, or the anisotropy of the interactions \cite{quian2013} lead to the breakdown or reduction of the blockade.

In this Letter, we show that, for experimentally relevant parameters, the Rydberg blockade is robust in ensembles of three atoms. In particular, we consider two different arrangements, namely, a line and an equilateral triangle. We observe an almost perfect van der Waals blockade and the coherent collective behavior of Rydberg excitations in both configurations. To go beyond this observation and understand  the dynamics of the system in detail, we measure the angular dependence of the effective interaction energy $V_{\rm eff}$ between two single-atoms excited to $\ket{r}\equiv\ket{nD_{3/2},m_j=3/2}$ Rydberg states. Using the measured two-body interaction strength we demonstrate that it is possible to fully reproduce the three-atom excitation dynamics in both the full and partial blockade regimes, with a model based on a master equation with no adjustable parameters. With the degree of experimental control demonstrated here, many theoretical proposals envisioning quantum simulation using Rydberg atoms become realistic.

\begin{figure}
\centering
\includegraphics[width=8.5cm]{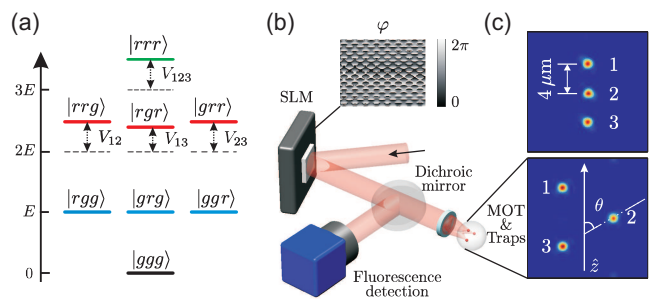}
\caption{(color online). (a) Relevant energy levels of a three-atom system with van der Waals interactions $V_{ij}$. In the blockade regime, the ground state $\ket{ggg}$ is resonantly coupled to the symmetric collective state $(\ket{ggr}+\ket{grg}+\ket{rgg})/\sqrt{3}$. (b) Scheme of the experimental setup. Arbitrary geometries of 2D arrays of dipole traps are obtained by imprinting a phase map $\varphi$ with the SLM. (c) Trap geometry. Three single-atoms are trapped in microscopic optical tweezers separated by~$R=4$ $\mu$m in a linear (top) and by ~$R=8$ $\mu$m in a triangular arrangement (bottom). The quantization axis $\hat{z}$ is set by a 3~G external magnetic field.}
\label{fig:fig1}
\end{figure}

We consider three atoms, with ground $\ket{g_i}$ and Rydberg $\ket{r_i}$ states coupled with Rabi frequencies $\Omega_i$, and interacting via pairwise interactions $V_{ij}$. The system is thus described by the Hamiltonian \cite{lesanovsky2011}
\begin{equation}
  \hat{H} = \sum_{i=1}^3 \frac{\hbar\Omega_i}{2}(\hat{\sigma}_{rg}^{(i)}+\hat{\sigma}_{gr}^{(i)}) + \sum_{i < j} V_{ij}\hat{\sigma}_{rr}^{(i)}\hat{\sigma}_{rr}^{(j)}
\label{eq_hamiltonian}
\end{equation}
where $\hat{\sigma}_{rg}^{(i)}=\ket{r_i}\langle g_i|$, $\hat{\sigma}_{gr}^{(i)}=\ket{g_i}\langle r_i|$, and $\hat{\sigma}_{rr}^{(i)}=\ket{r_i}\langle r_i|$. All parameters of the Hamiltonian can be tuned by a proper choice of the experimental settings. In particular, choosing $\ket{r}=\ket{nD_{3/2}}$ gives an extra degree of freedom to tune $V_{ij}$ due to the anisotropy of the interaction. In what follows all Rabi couplings $\Omega_i\equiv\Omega$ are equal within $5\%$.

A strong blockade is obtained if the interaction strengths $V_{ij}$ between atom pairs are much greater than the atom-light coupling $\hbar\Omega$ (Fig.~\ref{fig:fig1}(a)). In this regime, the states carrying double and triple Rydberg excitations are off-resonant with the light field and the system can be described as a two-level model involving the collective states $\ket{ggg}$ and $\ket{\Phi_{1r}}=(\ket{ggr}+\ket{grg}+\ket{rgg})/\sqrt{3}$, coupled by an effective Rabi frequency $\sqrt{3}\Omega$. Here, $\ket{ijk}\equiv\ket{i_1}\ket{j_2}\ket{k_3}$ stands for products of the single-atom ground $\ket{g}$, and Rydberg $\ket{r}$ states for atoms 1, 2, and 3.

Our apparatus, shown schematically in Fig.~\ref{fig:fig1}(b), was previously described in detail~\cite{beguin2013}. Three single $^{87}{\rm Rb}$ atoms are loaded from a magneto-optical trap into three 1~mK-deep microscopic optical traps~\cite{schlosser2001}, formed by focusing down a 850~nm gaussian beam to a waist of 1~$\mu$m ($1/e^2$ radius) using a high numerical aperture lens under vacuum~\cite{sortais2007}. Arbitrary patterns of traps are obtained by imprinting a calculated phase pattern on the beam with a spatial light modulator (SLM) \cite{nogrette2013}. CCD images of the two trap configurations used in this work are displayed in Fig.~\ref{fig:fig1}(c). In the first arrangement (top), the three traps are collinear (parallel to the quantization axis $\hat{z}$) and separated by $R= 4\,\mu {\rm m}$. In the second configuration (bottom), the traps form an equilateral triangle with 8 $\mu$m sides.

The same aspheric lens is used to collect the atom fluorescence from each trap. We trigger the experimental sequence as soon as one atom is detected in each of the three traps. The atoms are then optically pumped into $\ket{g}=\ket{5S_{1/2},F=2,m_F=2}$. The quantization axis $\hat{z}$ is set by a 3~G external magnetic field. For Rydberg excitation from $\ket{g}$ to $\ket{nD_{3/2},m_j=3/2}$, we use a two-photon process~\cite{miroshnychenko2010}: a $\pi$-polarized laser beam at 795~nm, detuned from the $\ket{5P_{1/2},F=2,m_F=2}$ intermediate state by $2\pi\times 740$~MHz, and a $\sigma^+$-polarized 474~nm laser beam. Both excitation lasers are frequency locked using an ultra-stable cavity providing laser linewidths $\sim 10$~kHz. During the Rydberg excitation, the dipole traps are switched off to avoid lightshifts. After excitation for a duration $\tau$, we switch on again the dipole traps and we look for the fluorescence of the three atoms. Excitation of an atom to the Rydberg state is inferred from its loss from the corresponding trap (and thus the absence of fluorescence), as Rydberg states are not trapped. The eight different populations $P_{ijk}$ of the three-atom states $\ket{ijk}$ are then reconstructed by repeating each sequence $\sim 150$ times~\cite{miroshnychenko2010}.

\begin{figure}
\centering
\includegraphics[width=8.5cm]{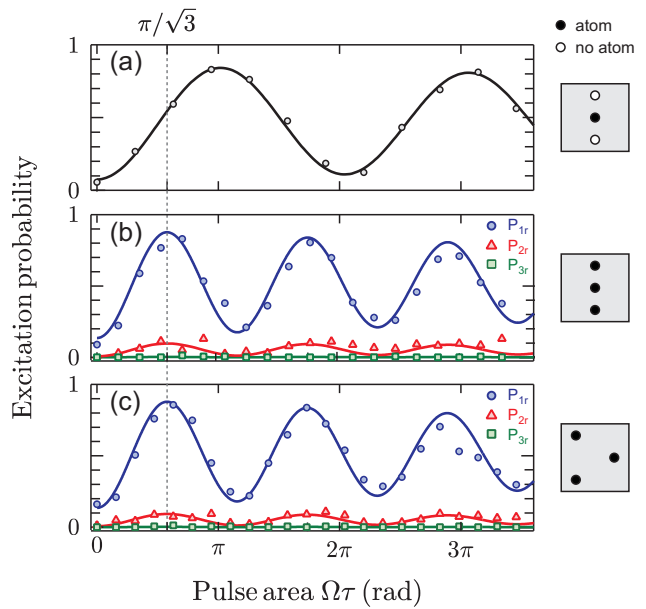}
\caption{(color online). (a) Representative single-atom Rabi flopping to the $\ket{82D_{3/2}}$ state for the central atom in the linear arrangement. Single-atom Rabi frequencies $\Omega \simeq 2\pi\times 0.8\,{\rm MHz}$, and damping rates $\gamma\simeq 0.3\,\mu {\rm s}^{-1}$ for all three atoms are obtained from fits (solid lines) to the solution of the OBEs for a single two-level atom. (b) Probability of single (blue circles), double (red triangles), and triple (green squares) Rydberg excitation as a function of the excitation pulse area in the linear arrangement. The collective enhancement of the Rabi frequency by $\sqrt{3}$ clearly appears in the data.  Solid lines are the result of the model described in the text without any adjustable parameter. (c) Same as (b) but for the triangular geometry.}
\label{fig:fig2}
\end{figure}

We first consider a 1D array of three individual atoms aligned along the quantization axis [see Fig.~\ref{fig:fig1}(c) top] and separated by $4\,\mu{\rm m}$. To obtain the single-atom Rabi frequencies $\Omega_i$ we measure the probability $P_{r_i}$ to excite atom $i$ to the Rydberg state, with the other two traps switched off, as a function of the excitation pulse area. We observe well-contrasted Rabi oscillations [Fig.~\ref{fig:fig2}(a)]. A fit of the data (solid line) gives the same Rabi frequencies $\Omega_i\simeq 2\pi\times 0.8\,{\rm MHz}$ for the three atoms (within $5\,\%$), as well as small damping rates $\gamma_i\simeq0.3\;{\rm \mu s}^{-1}$ (see below). In Fig.~\ref{fig:fig2}(b) a single atom is loaded in each of the three traps. In this configuration we expect full blockade, as the single-atom Rabi frequencies are much smaller than van der Waals interactions: even for the $R=8\,\mu{\rm m}$ distance between the outermost atoms, an extrapolation of the measurements of Ref.~\cite{beguin2013} give $V_{13}\simeq h\times 32\,{\rm MHz}$. The three atoms are excited to the collective state $\ket{\Phi_{1r}}$ [Fig.~\ref{fig:fig1}(a)], and the single excitation probability $P_{1r}\equiv P_{rgg}+P_{grg}+P_{ggr}$ shows oscillations with a frequency of $(1.72\pm 0.02)\Omega$, compatible with the expected $\sqrt{3}\Omega$. Clear blockade of multiple Rydberg excitations is observed in the data, as the populations $P_{2r} = P_{rrg}+P_{rgr}+P_{grr}$ (resp. $P_{3r}= P_{rrr}$) of doubly (resp. triply) excited states are almost totally suppressed in the system, with $P_{2r}$ (resp. $P_{3r}$) never exceeding $9\,\%$ (resp. $1\,\%$).

We now show that the actual blockade is even better than suggested by these values. Indeed, each atom has a small probability $\varepsilon$ to be lost during the sequence, independently of its internal state \cite{supplementary}. An independent measurement of the loss probability gives $\varepsilon = (5 \pm 1)\%$. Since in our detection scheme an atom loss is interpreted as an excitation to the Rydberg state, the \emph{observed} double excitation $P_{2r}$ differs from the \emph{actual} one $\tilde{P}_{2r}$ and, to first order in $\varepsilon$, it reads \cite{supplementary}
\begin{equation}
P_{2r} = (1-\varepsilon) \tilde{P}_{2r} + 2 \varepsilon \tilde{P}_{1r}.
\label{eq_prob}
\end{equation}
If the blockade were perfect, one would have $\tilde{P}_{2r}=0$, and the measured $P_{2r}$ would thus oscillate between 0 and $2\varepsilon$, in phase with $P_{1r}$. From the data on Fig.~\ref{fig:fig2}(b) we can extract an upper bound of $\sim 2\%$ on $\tilde{P}_{2r}$~\cite{supplementary}.

To gain more insight into the quality of the blockade for our experimental parameters, we simulate the dynamics of the system with Hamiltonian (\ref{eq_hamiltonian}). A sum of independent single atom dissipators
\begin{equation}
L[\rho]=\sum_i{\frac{\gamma_i}{2}(2\hat{\sigma}_{gr}^{(i)} \rho \hat{\sigma}_{rg}^{(i)}-\hat{\sigma}_{rr}^{(i)} \rho - \rho \hat{\sigma}_{rr}^{(i)})}.
\end{equation}
is used to account for a small experimental damping $\gamma_i$ of the oscillations (mainly due to off-resonant spontaneous emission through the intermediate state $\ket{5P_{1/2}}$; all the $\gamma_i$ are equal within 10~\%). The results of the simulation, with no adjustable parameter, are represented by solid lines in Fig.~\ref{fig:fig2}(b), where the loss-error correction (\ref{eq_prob}) is included. The very good agreement with the data further supports the quality of the blockade. Our results are compatible with the prediction of the model of double excitation probability $\tilde{P}_{2r}^{\rm (theo.)}\sim 10^{-3}$ for the same experimental parameters. Although proving experimentally that the double excitation is that low would require a more detailed study of systematic effects, this figure is very encouraging for high-fidelity generation of three-atom $\ket{\rm W}$ states.

In the results discussed so far, we only considered a 1D configuration. For scalability to a large number of atoms, however, 2D arrays of traps are preferable. In this case, some atom pairs necessarily have an internuclear axis not aligned along the quantization axis and the anisotropy of the interaction comes into play, which might eventually prevent a perfect blockade \cite{pohl2009,cano2012,quian2013}. To investigate this effect we study the blockade in an equilateral triangle configuration. Here, the anisotropic character of the $D$--state orbital plays a role and the interaction energies between atom pairs $V_{12}\simeq V_{23}$ are weaker than $V_{13}$, although the atoms are equally separated. Despite this, figure~\ref{fig:fig2}(c) shows that the strength of the blockade is not reduced in the triangular geometry. Double and triple excitation probabilities are inhibited and the single excitation probability oscillates at $\sim\sqrt{3}\Omega$. This result opens encouraging prospects for achieving strong blockade over 2D arrays of atoms.

In order to observe directly the anisotropy of the interaction~\cite{caroll2004} we measured the interaction energy between atom pairs separated by $R=12\,\mu{\rm m}$ as a function of the angle $\theta$ between the internuclear axis and the quantization axis $\hat{z}$. The procedure to extract the effective interaction energy $V_{\rm eff}$ is similar to the one introduced in Ref.~\cite{beguin2013}. Working in the partial blockade regime ($\hbar\Omega\sim V_{\rm eff}$), we model the excitation dynamics through the solution of the optical Bloch equations (OBE) involving two-level atoms. Strictly speaking, to model two atoms in the $\ket{nD_{3/2}}$ state and $\theta\neq 0$, one would need to consider all 49 Zeeman sublevels with their different van der Waals couplings~\cite{reinhard2007}. So as to keep the model tractable, even for large number of atoms, we model the system in the simplest non-trivial way, retaining only one single doubly excited state $\ket{rr}$ with an effective energy shift $V_{\rm eff}(\theta)$. All input parameters are obtained from single-atom Rabi oscillation experiments. The measured dynamics of the two-atom system are then fitted with the solution of the OBEs with $V_{\rm eff}$ as the only fitting parameter. A more detailed study of the angular dependence of the van der Waals interaction, taking into account the full Zeeman structure of the atom pair, is beyond the scope of this Letter and will be the subject of future work.

\begin{figure}
\centering
\includegraphics[width=8.5cm]{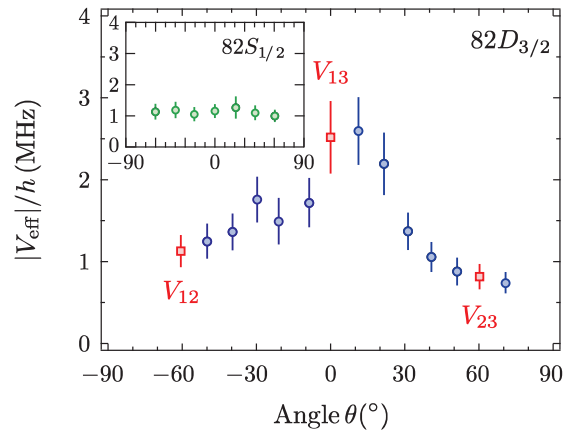}
\caption{(color online). Angular dependence of the effective interaction energy $V_{\rm eff}$ for two atoms in $\ket{82D_{3/2}}$  at $R=12\,\mu{\rm m}$, with $\theta$ the angle between the internuclear axis and the quantization axis $\hat{z}$. Red squares indicate the measured energy shifts $V_{13}$, $V_{23}$, and $V_{12}$ used for the simulation of three-atom dynamics in the partial blockade regime (see Fig.~\ref{fig:fig4}). In the inset, the angular dependence of the interaction for the spherically symmetric $\ket{82S_{1/2}}$ state is shown for comparison. Error bars represent one standard deviation confidence intervals in the fits.}
\label{fig:fig3}
\end{figure}

The result of this approach is shown in Fig.~\ref{fig:fig3} for the $\ket{82D_{3/2}}$ state. The anisotropy of the effective interaction is evident. The energy shift shows a maximum around $\theta=0$ and decreases for larger angles. A relative change of interaction strength by a factor $\sim 3$ is measured when $\theta$ varies from $\theta=0$ to $\theta=60^\circ$. In contrast, for a spherically symmetric $S$--Rydberg state, the interaction energy is isotropic (see inset of Fig.~\ref{fig:fig3})~\footnote{For excitation to the $S$-state, the polarisations of the 795~nm (resp. 474~nm) lasers are $\pi$ (resp $\sigma^-$).}. For the $D$-state, we observe an unexpected, slight asymmetry in the angular dependence of $V_{\rm eff}$, probably due to small systematic effects~\footnote{However, the observed asymmetry has a negligible impact on the three-atom dynamics that we study in the rest of the paper.}.  

\begin{figure}
\centering
\includegraphics[width=8.5cm]{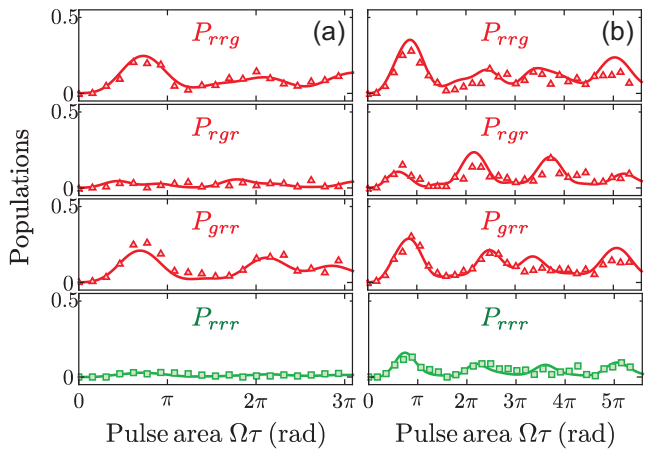}
\caption{(color online). Probabilities of detection of double Rydberg excitation $P_{rrg}$,$P_{rgr}$,$P_{grr}$, and triple excitation $P_{rrr}$ versus excitation pulse area $\Omega \tau$ for driving Rabi frequencies $\Omega=2\pi\times 0.8 \, {\rm MHz}$ (a), and $\Omega=2\pi\times$1.6 MHz (b) in the triangular configuration. The distance between the traps is $R=12 \mu$m. The ratio between effective pairwise interaction energies is $V_{13}/V_{12}\sim$ 3 for $\theta=$ 60$^\circ$. Solid lines are the solution of the OBEs without any adjustable parameter.}
\label{fig:fig4}
\end{figure}

The angular dependence of $V_{\rm eff}$ manifests itself in the interaction dynamics of the three atoms in the triangular configuration. By increasing the sides of the triangle to $R=12\,\mu{\rm m}$, the effective interaction energies become $(V_{12}, V_{23},V_{13})\simeq h \times (0.9, 1.1, 2.6) \, {\rm MHz}$ (see red squares in Fig.~\ref{fig:fig3}), and the blockade is only partial for our chosen Rabi frequency $\Omega$. In Fig.~\ref{fig:fig4} we show the populations of doubly ($P_{rrg}$, $P_{rgr}$, $P_{grr}$) and triply ($P_{rrr}$) excited states for two different Rabi frequencies. In the first dataset [Fig.~\ref{fig:fig4}(a)], $\Omega=2\pi\times 0.8 \, {\rm MHz}$ and the anisotropy in the binary interaction ($V_{12}\ne V_{13}$) is directly observed in the dynamics: the probability $P_{rgr}$ to detect double excitation of atoms 1 and 3 is almost totally suppressed, while it is appreciable for $P_{rrg}$ and $P_{grr}$. Those two curves show almost the same dynamics, as expected. Triple excitations are totally blocked in this regime. For comparison, we show also the dynamics when a slightly higher Rabi frequency $\Omega = 2 \pi\times 1.6\, {\rm MHz}$ is used [Fig.~\ref{fig:fig4}(b)]. This corresponds to a partial blockade regime where $V_{13}> \hbar\Omega > V_{12}$. In this case, even triple excitations are not completely blockaded. $P_{rrg}$ and $P_{grr}$ also exhibit similar behavior, while $P_{rgr}$ shows different dynamics. The populations of states carrying only single excitations also show the anisotropy~\cite{supplementary}.

Many-body effects have largely been recognized to play a key role in the modeling of systems in physics and chemistry~\cite{distasio2012}. In the case of Rydberg atoms they have been invoked to explain anomalous broadenings of F\"orster resonances~\cite{anderson1998,mourachko1998}. To understand the evolution of the population of the states during excitation and to investigate to what extent few-atom many-body physics can be described from pairwise interactions we perform again a simulation using the OBEs for the three-atom system. In the model, with no adjustable parameters, the measured interaction energies at $\theta=0$ and $\theta=60^\circ$ (red squares in Fig.~\ref{fig:fig3}) are introduced. As shown by the solid lines in Fig.~\ref{fig:fig4}, the simulation (where atom loss correction is included) fully reproduces the experimental data. The fact that the simulation can accurately describe the evolution of the triply excited state $P_{rrr}$ suggests that, for our choice of parameters, the pairwise addition of van der Waals level shifts $V_{123} = V_{12} + V_{13} + V_{23}$ is valid to a very good approximation. However, this additivity of the potential may not hold in the case of resonant dipole-dipole interactions. There, quantum interference between different many-body interaction channels can influence the dynamics~\cite{pohl2009}. All these processes can be studied for Rydberg atoms close to F\"orster resonance and will be the subject of future work. Another interesting line of research will consist in studying the recently predicted Borromean trimers bound by the dipole-dipole interaction~\cite{kiffner2013}.

In summary, we have investigated the dynamics of a system of three Rydberg atoms in both full and partial blockade regimes. We observe a strong van der Waals blockade of the excitations and coherent Rabi oscillations for two different spatial configurations. For the same experimental parameters in the equilateral triangle arrangement, the anisotropy of the interaction potential between $\ket{nD}$ states does not prevent the observation of a strong van der Waals blockade, which is a prerequisite for the scalability of quantum information processing proposals using 2D arrays of dipole traps. The strong blockade achieved and the small damping of the oscillations paves the way for the generation of many-atom entanglement with high fidelity through the Rydberg blockade~\cite{saffman2010}. In the partial blockade regime, the angular dependence of the interaction energy shift between two atoms has been measured for the $\ket{82D_{3/2}}$ and $\ket{82S_{1/2}}$ Rydberg states. Furthermore, we have shown that with the measured effective energy shifts it is possible to reproduce the three-atom dynamics with high accuracy. This result demonstrates that one can confidently scale those studies for 2D arrays of more than a few atoms, enabling the quantum simulation of large-size, long-range interacting spin systems.

We thank Daniel Cano for interesting discussions. This work was supported financially by the EU (ERC Stg Grant ARENA, AQUTE Integrating project, FET-Open Xtrack project HAIRS, EU Marie-Curie program ITN COHERENCE FP7-PEOPLE-2010-ITN-265031 (H.L.)), by the DGA (L.B.), and by R\'egion \^Ile-de-France (LUMAT and Triangle de la Physique, LAGON project).

\newpage

\onecolumngrid

\begin{center}
\large{\bf Supplemental Material: Demonstration of a strong Rydberg blockade \\ in three-atom systems with anisotropic interactions}
\end{center}

\author{D. Barredo}
\affiliation{Laboratoire Charles Fabry, Institut d'Optique, CNRS, Univ Paris Sud,\\
2 avenue Augustin Fresnel, 91127 Palaiseau cedex, France }

\author{S. Ravets}
\affiliation{Laboratoire Charles Fabry, Institut d'Optique, CNRS, Univ Paris Sud,\\
2 avenue Augustin Fresnel, 91127 Palaiseau cedex, France }

\author{H. Labuhn}
\affiliation{Laboratoire Charles Fabry, Institut d'Optique, CNRS, Univ Paris Sud,\\
2 avenue Augustin Fresnel, 91127 Palaiseau cedex, France }

\author{L. B\'eguin}
\affiliation{Laboratoire Charles Fabry, Institut d'Optique, CNRS, Univ Paris Sud,\\
2 avenue Augustin Fresnel, 91127 Palaiseau cedex, France }

\author{A. Vernier}
\affiliation{Laboratoire Charles Fabry, Institut d'Optique, CNRS, Univ Paris Sud,\\
2 avenue Augustin Fresnel, 91127 Palaiseau cedex, France }

\author{F. Nogrette}
\affiliation{Laboratoire Charles Fabry, Institut d'Optique, CNRS, Univ Paris Sud,\\
2 avenue Augustin Fresnel, 91127 Palaiseau cedex, France }

\author{T. Lahaye}
\affiliation{Laboratoire Charles Fabry, Institut d'Optique, CNRS, Univ Paris Sud,\\
2 avenue Augustin Fresnel, 91127 Palaiseau cedex, France }

\author{A. Browaeys}
\affiliation{Laboratoire Charles Fabry, Institut d'Optique, CNRS, Univ Paris Sud,\\
2 avenue Augustin Fresnel, 91127 Palaiseau cedex, France }

\maketitle

\section{Full dataset for the three-atom dynamics in the partial blockade regime}

\begin{figure*}[b!]
\centering
\includegraphics[width=17cm]{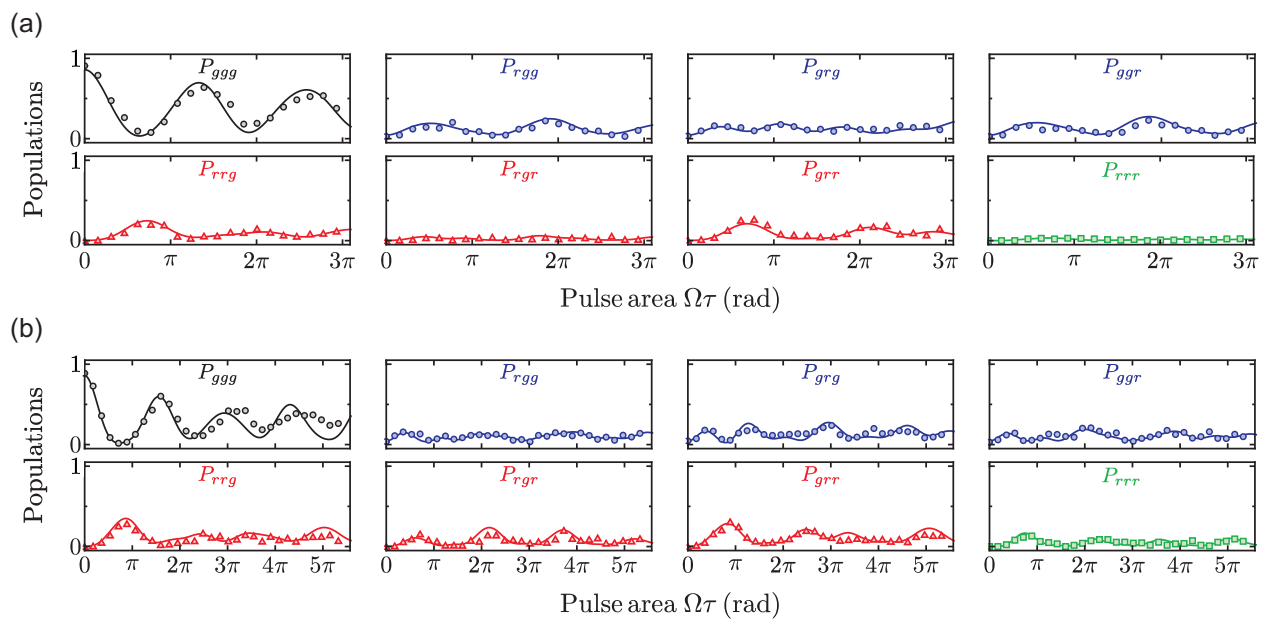}
\caption{Populations $P_{ijk}$ of the three-atom system versus excitation pulse area $\Omega \tau$ for driving Rabi frequencies $\Omega_i\simeq2\pi\times 0.8 \, {\rm MHz}$ (a), and $\Omega_i\simeq2\pi\times$1.6 MHz (b). The distance between the traps is $R=12 \mu$m. The ratio between effective pairwise interaction energies is $V_{13}/V_{12}\sim$ 3 for $\theta=$ 60$^\circ$. Solid lines are the solution of the OBEs for the three-atom system without any adjustable parameter.}
\label{fig:fig_sup}
\end{figure*}

The dynamics of the three-atom system was measured for two different Rabi frequencies. In Fig. S\ref{fig:fig_sup} we show, together with the data, the results of the simulation of the partial blockade data for all the state populations (see Fig. 4 of the main text). The model includes the loss-error correction discussed in details below. In both datasets the anisotropy of the interaction is also clearly visible when comparing the states carrying single excitations: the probability $P_{grg}$ displays different dynamics when compared to the populations $P_{rgg}$ and $P_{ggr}$. The overall agreement with the simulations is very good, further supporting the high degree of control of the three-atom system.

\section{Accounting for loss errors in the model}

In order to directly compare the observed data in the full blockade regime (Fig. 2 of the main text) with our simulations we need to account for small imperfections in the experiment. In our detection method, the successful excitation of an atom to a Rydberg state is inferred from the absence of fluorescence at the corresponding trap sites. Each atom has a small probability $\varepsilon$ to be lost during the sequence, independently of its internal state. Losses come from two effects: (i) collisions with background gas, and (ii) escape from the trap region during the Rydberg excitation sequence due to the finite temperature of the atoms. Both contributions add up to an $\varepsilon\simeq 5\%$ loss probability for the parameters of our sequence. When this happens the event is erroneously counted as a Rydberg excitation, thus reducing the apparent blockade. In the case of a single atom, the observed probability to be in the ground state $P_g$ is therefore $P_g= (1-\varepsilon)\tilde{P}_g$, where $\tilde{P}_g$ is the actual probability for the atom to be in the ground state. On the contrary, the observed probability for Rydberg excitation $P_r$ is related to the actual value $\tilde{P}_r$ by $P_r=(1-\varepsilon)\tilde{P}_r + \varepsilon$.
Generalized to the case of three atoms the relation between the observed probabilities $P_{ijk}$ and the actual ones $\tilde{P}_{ijk}$ reads:
\begin{align}
P_{ggg} &= (1-\varepsilon)^3 \tilde{P}_{ggg}\\
P_{rgg} &= (1-\varepsilon)^2 (\tilde{P}_{rgg}+ \varepsilon \tilde{P}_{ggg})\\
P_{grg} &= (1-\varepsilon)^2 (\tilde{P}_{grg}+ \varepsilon \tilde{P}_{ggg})\\
P_{ggr} &= (1-\varepsilon)^2 (\tilde{P}_{ggr}+ \varepsilon \tilde{P}_{ggg})\\
P_{rrg} &= (1-\varepsilon) [\tilde{P}_{rrg}+ \varepsilon ( \tilde{P}_{rgg}+ \tilde{P}_{grg}) + \varepsilon^2 \tilde{P}_{ggg}]\\
P_{rgr} &= (1-\varepsilon) [\tilde{P}_{rgr}+ \varepsilon ( \tilde{P}_{ggr}+ \tilde{P}_{rgg}) + \varepsilon^2 \tilde{P}_{ggg}]\\
P_{grr} &= (1-\varepsilon) [\tilde{P}_{grr}+ \varepsilon ( \tilde{P}_{grg}+ \tilde{P}_{ggr}) + \varepsilon^2 \tilde{P}_{ggg}]\\
P_{rrr} &= \tilde{P}_{rrr}+ \varepsilon (\tilde{P}_{grr}+ \tilde{P}_{rgr}+ \tilde{P}_{rrg})+  \varepsilon^2 (\tilde{P}_{ggr}+\tilde{P}_{rgg}+\tilde{P}_{grg}) + \varepsilon^3 \tilde{P}_{ggg}.
\end{align}
Here, no attempt was made to correct the data for other sources of imperfections like inefficient optical pumping to the $\ket{5S_{1/2},F=2,m_F=2}$ state or finite ionization and decay rates of the Rydberg state~[S1].

Solving the equations for $\tilde{P}_{2r}$ ($\tilde{P}_{2r}=\tilde{P}_{grr}+\tilde{P}_{rgr}+\tilde{P}_{rrg}$) and keeping only terms up to order $\varepsilon$, yields $\tilde{P}_{2r}=P_{2r}+(P_{2r}-2 P_{1r })\varepsilon + O(\varepsilon^2)$. Through Monte Carlo simulations where we assume that $P_{2r}$, $P_{1r}$ and $\varepsilon$ are normally distributed random variables (with means and standard deviations given by fitting the experimental data), we estimate the actual double excitation probability to be $\tilde{P}_{2r}\simeq 0.01\pm 0.01$, which is in agreement with the double excitation probability $\tilde{P}_{2r}^{\rm (theo.)}\sim 10^{-3}$ predicted by the model. We thus conservatively put an experimental upper bound of 2~\% for the double excitation probability.


\begin{thebibliography}{30}

\bibitem{nielsen2000}
M. A. Nielsen, I. L. Chuang, Quantum computation and quantum information (Cambridge University Press, Cambridge) (2000).

\bibitem{feynman1982}
R. Feynman, Int. J. Theor. Phys. {\bf 21}, 467-488 (1982).

\bibitem{lanyon2011}
B. P. Lanyon \emph{et al.}, Science {\bf 334}, 57-61 (2011).

\bibitem{britton2012}
J. W. Britton \emph{et al.}, Nature {\bf 484}, 489-492 (2012).

\bibitem{islam2013}
R. Islam \emph{et al.}, Science {\bf 340}, 583-587 (2013).

\bibitem{weimer2010}
H. Weimer, M. M\"uller, I. Lesanovsky, P. Zoller and H. P. B\"uchler, Nat. Phys. {\bf 6}, 382 (2010).

\bibitem{jaksch2000}
D. Jaksch \emph{et al.}, Phys. Rev. Lett. {\bf 85}, 2208 (2000).

\bibitem{lukin2001}
M. D. Lukin \emph{et al.}, Phys. Rev. Lett. {\bf 87}, 037901 (2001).

\bibitem{comparat2010}
D. Comparat and P. Pillet, J. Opt. Soc. Am. B {\bf 27}, A208 (2010).

\bibitem{wilk2010}
T. Wilk \emph{et al.}, Phys. Rev. Lett. {\bf 104}, 010502 (2010).

\bibitem{isenhower2010}
L. Isenhower \emph{et al.}, Phys. Rev. Lett. {\bf 104}, 010503 (2010).

\bibitem{saffman2010}
M. Saffman, T. G. Walker, and K. M{\o}lmer, Rev. Mod. Phys. {\bf 82}, 2313 (2010).

\bibitem{singer2004}
K. Singer, M. Reetz-Lamour, T. Amthor, L. G. Marcassa, and M. Weidem\"uller, Phys. Rev. Lett. {\bf 93} 163001 (2004).

\bibitem{heidemann2007}
R. Heidemann \emph{et al.}, Phys. Rev. Lett. {\bf 99}, 163601 (2007).

\bibitem{pritchard2010}
J. D. Pritchard \emph{et al.}, Phys. Rev. Lett. {\bf 105}, 193603 (2010).

\bibitem{dudin2012}
Y. O. Dudin and A. Kuzmich, Science {\bf 336}, 887 (2012).

\bibitem{dudin2012b}
Y. O. Dudin, L. Li, F. Bariani, and A. Kuzmich, Nat. Phys. {\bf 8}, 790 (2012).


\bibitem{schauss2012}
P. Schauss \emph{et al.}, Nature {\bf 491}, 87 (2012).

\bibitem{urban2009}
E. Urban \emph{et al.}, Nat. Phys. {\bf 5}, 110 (2009).

\bibitem{gaetan2009}
A. Ga\"etan \emph{et al.}, Nat. Phys. {\bf 5}, 115 (2009).

\bibitem{gunter2013}
G. G\"unter \emph{et al.}, Science {\bf 342}, 954 (2013).

\bibitem{bidermann2014}
A. M. Hankin \emph{et al.}, arXiv:1401.2191 (2014).

\bibitem{pohl2009}
T. Pohl, and P. R. Berman, Phys. Rev. Lett. {\bf 102}, 013004 (2009).

\bibitem{cano2012}
D. Cano and J. Fort\'agh, Phys. Rev. A {\bf 86}, 043422 (2012).

\bibitem{quian2013}
J. Quian, X-D Zhao, L. Zhou, and W. Zhang, Phys. Rev. A {\bf 88}, 033422 (2013).

\bibitem{lesanovsky2011}
I. Lesanovsky, Phys. Rev. Lett {\bf 106}, 025301 (2011).

\bibitem{beguin2013}
L. B\'eguin \emph{et al.}, Phys. Rev. Lett. {\bf 110}, 263201 (2013).

\bibitem{schlosser2001}
N. Schlosser, G. Reymond, I. Protsenko and P. Grangier, Nature {\bf 411}, 1024 (2001).

\bibitem{sortais2007}
Y. R. P. Sortais \emph{et al.}, Phys. Rev. A {\bf 75}, 013406 (2007).

\bibitem{nogrette2013}
F. Nogrette \emph{et al.}, to be published.

\bibitem{miroshnychenko2010}
Y. Miroshnychenko \emph{et al.}, Phys. Rev. A {\bf 82}, 013405 (2010).

\bibitem{caroll2004}
T. J. Caroll \emph{et al.}, Phys. Rev. Lett. {\bf 93}, 153001 (2004).

\bibitem{supplementary}
See Supplemental Material at [URL will be inserted by publisher] for detailed error-loss analysis and full system dynamics.

\bibitem{reinhard2007}
A. Reinhard \emph{et al.}, Phys. Rev. A {\bf 75}, 032712 (2007).

\bibitem{distasio2012}
R. A. DiStasio Jr. \emph{et al.}, PNAS {\bf 109}, 14791 (2012).

\bibitem{anderson1998}
W. R. Anderson \emph{et al.}, Phys. Rev. Lett. {\bf 80}, 249 (1998).

\bibitem{mourachko1998}
I. Mourachko \emph{et al.}, Phys. Rev. Lett. {\bf 80}, 253 (1998).

\bibitem{kiffner2013}
M. Kiffner, W. Li, and D. Jaksch, Phys. Rev. Lett. {\bf 111}, 233003 (2013).

\end{thebibliography}

\begin{thebibliography}{30}


\bibitem{urban2009}
E. Urban \emph{et al.}, Nat. Phys. {\bf 5}, 110 (2009).


\end{thebibliography}
\end{document}